# Compressive Sensing for Feedback Reduction in MIMO Broadcast Channels

Syed T. Qaseem and Tareq Y. Al-Naffouri

## Abstract

We propose a generalized feedback model and compressive sensing based opportunistic feedback schemes for feedback resource reduction in MIMO Broadcast Channels under the assumption that both uplink and downlink channels undergo block Rayleigh fading. Feedback resources are shared and are opportunistically accessed by users who are *strong*, i.e. users whose channel quality information is above a certain fixed threshold. Strong users send same feedback information on all shared channels. They are identified by the base station via compressive sensing. Both analog and digital feedbacks are considered. The proposed analog & digital opportunistic feedback schemes are shown to achieve the same sum-rate throughput as that achieved by dedicated feedback schemes, but with feedback channels growing only logarithmically with number of users. Moreover, there is also a reduction in the feedback load. In the analog feedback case, we show that the propose scheme reduces the feedback noise which eventually results in better throughput, whereas in the digital feedback case the proposed scheme in a noisy scenario achieves almost the throughput obtained in a noiseless dedicated feedback scenario. We also show that for a fixed given budget of feedback bits, there exist a trade-off between the number of shared channels and thresholds accuracy of the feedback SINR.

## I. Introduction

Recently, it has been shown that dirty paper coding (DPC) achieves the sum-rate throughput of the multiple-input multiple-output (MIMO) broadcast channel [1], [2]. However, it requires a great deal of feedback as the transmitter needs perfect channel state information for all users and is computationally expensive [3]. Since then, many works have attempted to achieve the same sum-rate throughput with imperfect channel state information (reduced feedback load). This was done by applying opportunistic communication in the forward link [4]- [7].

By reviewing the feedback protocols suggested in literature, one can note that generally the following three components are fed back and user selection is based on either one or a combination of these components [4]- [9]:

1) Channel Direction Information (CDI), e.g., beam index(BI), quantized channel index (QCI)... etc.

2) Channel Quality Information (CQI), e.g., SNR, SINR, channel norm etc.

3) User identity (ID)

Feedback schemes can be differentiated according to whether the feedback is analog or digital. It is termed digital if the feedback involves only digital data (integer or bits) and analog otherwise. Feedback schemes can also be classified as either opportunistic or non-opportunistic.

While the forward link is opportunistic in nature, almost all feedback schemes are non-opportunistic. Here, each user has a dedicated feedback channel. For example, in random beamforming (RBF) scheme proposed by Sharif and Hassibi [4], the users are differentiated according to the channel direction (i.e. what beam direction is the channel mostly aligned with) and accordingly users feedback to Base Station (BS) the SINR corresponding to that direction *only*. So, each user feedback one integer and one real number. In order to reduce the feedback load further Diaz et. al. in [7] propose a threshold based RBF. Here, instead of feeding back the SINR for the best beam for each receive antenna (one real plus one integer numbers), the user only transmits one bit to the BS, indicating whether or not the SINR on a pre-selected beam for any receive antenna is above a given threshold. The scheme is repeated for each beam. Since interaction or cooperation among the competing users is not allowed, hence defying opportunism, there is a linear increase in the feedback resources (or channels) with the number of users [5], [6]. Even if thresholding is applied, there is no reduction in the number of feedback channels. This is because the channels are reserved even when users are not sending any feedback information.

Recently, some works have started to consider opportunistic feedback schemes where feedback resources are shared and are opportunistically accessed by strong users i.e. users whose CQI is above the given thresholds. Thus, in [8], Tang et. al. propose a feedback scheme with fixed number of feedback slots (channels) that are randomly accessed by strong users. In every slot, each strong user independently attempts to send back to the BS a data package containing its user identity (ID) with a probability. If two or more users feedback in the same slot, collision occurs and the feedback in that slot is discarded. In the case when multiple users successfully feeding back, the BS randomly selects one of the successful users. Although the scheme requires only an integer feedback per slot, it is suboptimal as the user is selected randomly. The scheme was only proposed for the single-input single-output (SISO) case. In [9], Rajiv et. al. propose a feedback scheme based on random access slots for the MIMO case that requires only user identity feedback (also an integer feedback per slot). In this scheme, time is divided into slots that are equally divided among the beams (CDI). Each slot then corresponds to a pre-determined threshold. Thus, if a user's CQI (e.g. SINR) on a particular beam exceeds the threshold corresponding to a particular slot, that user would feedback on that slot.

Both of the two schemes above require accurate timing-synchronization to avoid collisions, which is difficult to achieve in practice. Moreover, feedback to the BS is successful if there are no collisions, i.e., only one user is attempting to feed back in a slot. In addition, the two schemes only work for digital feedback but not when the designer is interested in analog feedback. In all the feedback schemes discussed above, the feedback links were assumed ideal when the forward links were subject to both fading and noise. This asymmetry in the way the two links are treated is unrealistic.

In this paper, we consider a broadcast scenario where the forward and the feedback links are symmetric in that they are both i) non-ideal and ii) opportunistic or shared. Thus, both links undergo Rayleigh fading and are subject to additive Gaussian noise. Moreover, the channels in both links are shared and are opportunistic in the sense feedback channels are dominated by strong users. Finally, the feedback links can be used for both analog and digital feedback.

The paper proposes a generalized feedback model and compressive sensing (CS) [10]- [14] based opportunistic feedback protocols for feedback resource reduction. Just as in all existing feedback techniques, a number of channel directions or beam is first determined. For each direction, the number of feedback channels is fixed and strong users fedback their CQI information on all feedback channels. In the analog feedback case, each strong user feds back CQI value whereas in the digital feedback case, each strong user feds back "1" if his CQI is above a particular threshold and remains silent otherwise. This creates an undetermined system of equations in a sparse vector of users. We use the emerging compressive sensing technique to identify users who have fed back and to estimate the fedback CQI. Users with higher value CQI have a stronger chance of being recovered. The results obtained via compressive sensing are refined using least-squares. As the feedback links are noisy, so the BS backs off on the noisy CQI based on the variance of the noise. We obtain the optimum back off on the noisy CQI that maximizes the throughput. A user among strong users is selected (strongest in the analog feedback case & randomly in the digital feedback case). The scheme is repeated for each channel direction. Although we have used SINR feedback, the proposed schemes can work with any kind of CQI (e.g. SNR). It is important to note that our scheme is less sensitive to timing-synchronization errors, as the scheme will be affected only if out of synchronization user is selected for a particular channel direction (the probability of which is low).

The remainder of the paper is organized as follows. In section II, generalized feedback model is introduced. In section III we discuss the proposed feedback strategy. In section IV, we present the sum-rate throughput obtained by the proposed schemes in the RBF case. In section V, performance evaluation of the proposed feedback schemes is presented. Feedback channel training is discussed in section VI followed by numerical results and conclusions in sections VII and VIII respectively.

*Notation*: We use bold upper and lower case letters for matrices and vectors, respectively. $\mathbf{A}^T$, $\mathbf{A}^*$ and $\mathbf{A}^\dagger$ refers to Transpose, Hermitian conjugate and pseudo-inverse of $\mathbf{A}$ respectively. $\mathbb{E}_{\mathbf{x}}[\cdot]$ denotes the expectation operator w.r.t. $\mathbf{x}$, and $\mathbb{P}[\ ]$ is the probability of the given event. The natural logarithm is referred to as $\log(\cdot)$, while the base 2 logarithm is denoted as $\log_2(\cdot)$. $f(x) = O(g(x))$ is equivalent to $f(x) = cg(x)$ where $c$ is a constant. $|A|$ denotes the size of a set $A$.

## II. System Model

### A. Downlink Transmission Model

We consider a single cell multi-antenna broadcast channel with $p$ antennas at the base station (transmitter) and $n$ users (receivers) each having one antenna. The channel is described by a propagation matrix which is constant during the coherence interval and is known completely at the receiver. Let $\mathbf{u} \in \mathbb{C}^{p \times 1}$ be the transmit symbol vector and let $x_i$ be the received signal by the $i$-th user, the received signal by the $i$-th user can then be written as

$$x_i = \sqrt{\rho_i}\mathbf{h}_i\mathbf{u} + w_i, \qquad\qquad i = 1, \ldots, n \qquad\qquad (1)$$

where $\mathbf{h}_i \in \mathbb{C}^{1 \times p}$ is the channel gain vector between the transmitter and the user, and $w_i$ is the additive noise. The entries of $\mathbf{h}_i$ and $w_i$ are i.i.d. complex Gaussian with zero mean and unit variance, $\mathcal{CN}(0, 1)$. Moreover, $\mathbf{u}$ satisfies an average transmit power constraint $\mathbb{E}\{\mathbf{u}^*\mathbf{u}\} = 1$ and $\rho_i$ is the SNR of the $i$-th user. A homogeneous network is considered, in which all users have the same SNR, i.e. $\rho_i = \rho = P/p$ for $i = 1, \ldots, n$, where $P$ is the total power available at the transmitter assuming that the noise power is unity. We also assume that the number of mobiles is greater than or equal to the number of transmit antennas, i.e., $n \geq p$, and that the BS selects $p$ out of $n$ users to transmit to.

### B. Generic Multi-antenna Feedback Channel

We present here a general model for the multiuser feedback channel with $r$ feedback channels (possibly shared) among $n$ users, in which users report channel quality information (CQI) to the base station in order to exploit multiuser diversity. As we shall soon see, this model encompasses the existing feedback models. The feedback channels are described by a propagation matrix $\mathbf{A}$ which is constant during the coherence interval and is assumed to be perfectly known at the BS (receiver), and are to be independent of the downlink channel. Let $\mathbf{v} \in \mathbb{C}^{n \times 1}$ be transmit feedback vector and let $y_i$ be the signal received via the $i$-th feedback channel. The signal received through the $i$-th feedback channel is mathematically described as

$$
\begin{bmatrix} y_1 \\ y_2 \\ \vdots \\ y_r \end{bmatrix} = \begin{bmatrix} a_{11} & a_{12} & \cdots & a_{1n} \\ a_{21} & a_{22} & \cdots & a_{2n} \\ \vdots & \vdots & \vdots & \vdots \\ a_{r1} & a_{r2} & \cdots & a_{rn} \end{bmatrix} \begin{bmatrix} v_1 \\ v_2 \\ \vdots \\ v_n \end{bmatrix} + \begin{bmatrix} w_1 \\ w_2 \\ \vdots \\ w_r \end{bmatrix}
$$

or equivalently

$$\mathbf{y} = \mathbf{A}\mathbf{v} + \mathbf{w} \tag{2}$$

where $r \leq n$ and $a_{ij}$ represents the (generally complex valued) gain of the $i$-th channel for the $j$-th user. Note that in contrast to the majority of existing feedback reduction techniques, a noisy feedback channel is assumed. The entries of $\mathbf{w}$ represent the additive noise and are assumed to be i.i.d. complex Gaussian with zero mean and variance $\sigma^2$, $\mathcal{CN}\left(0, \sigma^2\right)$.

If no fading is considered ($\mathbf{A}$ is deterministic), all entries of $\mathbf{A}$ are equal to a constant, whereas if the feedback channels undergo block Rayleigh fading, $\mathbf{A}$ remains constant during the coherence interval and its entries are i.i.d. complex Gaussian with zero mean and unit variance, $a_{ij} \sim \mathcal{CN}\left(0, 1\right)$, assumed to be known perfectly at the BS via feedback channel training (discussed in Section VI).

We summarize in Table I, how our feedback model (2) applies to the feedback models (opportunistic or not) suggested in literature [4]- [9]. Thus, in the non-opportunistic feedback model, each user is allocated its own feedback channel and the uplink channel matrix $\mathbf{A}$ becomes diagonal and of size $n$ (equal to the number of users). For the opportunistic models proposed in [8] by Tang et. al., the feedback channel matrix $\mathbf{A}$ becomes diagonal of size $r \times r$, where $r$ is the number of feedback slots and is less than $n$. $\mathbf{v}$ represents feedback data in each slot, and when a collision in a particular slot takes place, the corresponding entry of $\mathbf{v}$ is not valid. The same model holds for [9] except that in this scheme $r$ is not fixed but varies randomly. Also, $r$ may not necessarily be less than $n$. In all these schemes, the additive noise $\mathbf{w}$ is set to zero.

In this paper, we take a more general approach and consider a contention-based feedback protocol, which assigns independent multi-access contention channels for CQI reporting in which the feedback process should itself be a filter that selects strongest users. There are different ways to interpret the system of equations (cf. (2)). One possibility is to assume that each user is equipped with one antenna and the BS is equipped with $r$ antennas. In this case $a_{ij}$ represents the gain from the $j$-th user to the $i$-th antenna (spatial feedback channels). Another possibility is to assume that each single-antenna user is going to feedback the same information over $r$ frequency bands shared with the other users. Thus, $a_{ij}$ represents the gain of the $j$-th user in the $i$-th band (frequency feedback channels).

## III. PROPOSED FEEDBACK STRATEGY

Before we discuss the proposed feedback strategy, we present important compressive sensing results used in our work. A short introduction to compressive sensing is given in the Appendix.

### A. Sparsity Pattern Recovery Results

Compressive sensing refers to the recovery of the sparsity pattern $S$ (with $|S| = s$) of signal $\mathbf{v} \in \mathbb{R}^n$ accurately from limited measurements

$$S = \{i \in \{1, \ldots, n\} | v_i \neq 0\} \tag{3}$$

Two approaches for recovering the sparsity pattern in the noisy setting (cf. (2)) are discussed here, the only exception being that these results are derived for the case when the entries of $\mathbf{A}$ and $\mathbf{w}$ are i.i.d. real Gaussians i.e., $a_{ij} \sim \mathcal{N}(0, 1)$ and $w_i \sim \mathcal{N}(0, \sigma^2)$.

*1) Sparsity pattern recovery using LASSO [13]:* A recent paper by Wainwright [13] shows that it is surprisingly possible to recover the sparsity pattern of signals accurately from limited measurements in a noisy setting using LASSO which is $l_1$-constrained quadratic program (QP). The LASSO gives the estimated $\hat{\mathbf{v}}$.

$$\hat{\mathbf{v}} = \arg \min_{\mathbf{v} \in \mathbb{R}^{n \times 1}} \left\{ \frac{1}{2r_1} \|\mathbf{y} - \mathbf{A}\mathbf{v}\|_2^2 + \alpha \|\mathbf{v}\|_1 \right\} \tag{4}$$

where $\alpha > 0$ is a user-defined regularization parameter.

The number of measurements (or channels) required for successful sparsity pattern recovery using LASSO must satisfy $r_1 = c_1 s \log(n - s)$, where $c_1$ is a constant.

*2) Sparsity pattern recovery using Maximum Correlation [14]:* Similar results for sparsity pattern recovery from limited measurements in a noisy setting using maximum correlation, a much simpler method compared to LASSO, are derived by Fletcher et. al. in [14]. Maximum Correlation estimate is defined as the indices corresponding to the $s$ largest values of $\hat{I}$, where $\hat{I}$ is defined as follows

$$\hat{I} = \mathbf{A}^T \mathbf{y}$$

                                                                                

The number of measurements (or channels) required for successful sparsity pattern recovery using maximum correlation must satisfy $r_2 = c_2 s \log(n - s)$, where $c_2$ is a constant.

*3) Refining CS results using Least Squares:* In our framework, we propose to refine the results obtained via compressive sensing through least squares (LS) using the following procedure: once the sparsity pattern $S$ is known, we can form matrix $\mathbf{A}_S$ with columns of $\mathbf{A}$ corresponding to $S$ and estimate $\mathbf{v}$ as $\mathbf{v}_{LS} = \mathbf{A}_S^\dagger \mathbf{y}$

### B. General Strategy of Using Compressive Sensing for Feedback

Any feedback scheme has two components, a direction component and a magnitude component. The transmitter usually has certain pre-determined directions for which it seeks user feedback. Thus, the BS announces that it is seeking feedback for a particular direction. At this instant, the users whose channels lie at or are close to this direction, feedback their CQI (SNR, SINR, channel strength etc.). Now a limited number of users will feedback on the set of shared feedback channels according to input/output equation (2).

Thus the vector $\mathbf{v}$ in (2) is sparse with sparsity level determined by the number of users who feedback. CS can now be used to recover the sparsity pattern of $\mathbf{v}$ [13]- [14] (i.e. which user prefer that particular direction) and could also recover the vector $\mathbf{v}$ itself [13]) (i.e. users' feedback CQI). Moreover, the larger the value of particular CQI, the higher the chances of its recovery. Another factor that enhances the level of recovery is how sparse the vector $\mathbf{v}$ as compared to the number of feedback channels available. We need at least one strong user (i.e. $s \geq 1$) for each beam or direction in order to achieve full multiplexing gain which implies that small values of $s$ are sufficient. To reduce the number of users who feedback $s$, we pursue a thresholding strategy where the user will feedback if his CQI is greater than a threshold $\zeta$ to be determined.

Now consider a particular beam (CDI) (all beams will behave in an identical manner as the users are i.i.d. and the beams are equi-powered). Noting that the users' CQI are i.i.d., we can choose $\zeta$ to produce a sparsity level $s$. This happen by requiring that

$$\bar{F}(\zeta) = \arg \max_{u \in (0,1)} \binom{n}{s} u^s (1-u)^{n-s} \tag{5}$$

where $\bar{F}(\zeta)$ or $u$ is the complementary cumulative distribution function (CCDF) of CQI (SINR) defined as: $\bar{F}(\zeta) = \mathbb{P}[\text{SINR} > \zeta] = \frac{\exp(-\zeta/\rho)}{(1+\zeta)^{p-1}}, \zeta \geq 0$.

*Lemma 1: Threshold that maximizes (5) is given by $\zeta = \bar{F}^{-1}\left(\frac{s}{n}\right)$*

Proof: Let $\psi = \binom{n}{s} u^s (1-u)^{n-s}$. Differentiating $\psi$ w.r.t $u$ and setting the derivative to 0, and solving for $u$ yields $u = s/n$. Thus, $\bar{F}(\zeta) = s/n$, or $\zeta = \bar{F}^{-1}(s/n)$.

*C. Feedback Protocol for the Analog Feedback Case*

In the analog feedback scenario, users above threshold feed back their analog CQI value. The CS strategy then allows the BS to recover all users who transmitted their CQI. This off course will be true provided that the number of users who feedback is less than or equal to $s$. Here, we assume that the probability the a user is strongest for more than one beam is negligible as the number of users are relatively much larger than the number of beams. It has been shown in [4] that this is a valid assumption under these conditions. The steps of the proposed compressive sensing based opportunistic feedback protocol are as follows:

1) **Threshold Determination:** BS decides on thresholding level $\zeta$ based on the sparsity level that can be recovered.

2) **User Feedback:** Repeat the following steps for each beam.

   - CQI Determination: Each user determines his best beam (corresponding to the highest CQI value).
   - CQI Feedback: Each user feeds back his CQI if it is higher than $\zeta$ on all shared channels. Otherwise, the user remains silent.
   - Compressive Sensing: BS finds the strong users using CS.
   - Least-squares estimation/refining: BS estimates or refines results obtained via CS using least-squares.
   - Optimum CQI Back off: BS backs off on the noisy CQI (SINR) based on the noise variance such that the throughput is maximized.

3) **User Selection:** Select users and schedule them to beams.

*Remark 1:* Once CQI has been determined for each beam direction, the base station can proceed to implement any of the various multiuser scheduling techniques. For example, the BS can go for random beamforming and might opt for a second-stage feedback to design the final precoding matrix [6]. The second-stage feedback is requested from the selected users only. Thus, the amount of second stage feedback is relatively much smaller compared to the amount of first-stage feedback. Alternatively, the BS could also implement the semi-orthogonal user selection (SUS) algorithm (zero beamforming) proposed in [5].

*D. Feedback Protocol for the Digital Feedback Case*

The digital feedback is similar to analog feedback except that user feeds back "1" if his CQI for a particular beam is above a particular threshold. Otherwise, the user remains silent. To increase the feedback granularity, we let the users compare his CQI to a set of thresholds, not just one. Thus, suppose that we want to set $k$ thresholds $\zeta_1 < \zeta_2 < \ldots , < \zeta_k$ such that the number of users whose CQI lie between the two consecutive thresholds $[\zeta_i, \zeta_{i+1})$ is equal to $s$. Note that the last interval is $[\zeta_k, \infty)$. Following our discussion in subsection III-B, we can set the lowermost threshold as

$$\bar{F}(\zeta_1)n = sk, \quad \text{or,} \quad \zeta_1 = \bar{F}^{-1}\left(\frac{sk}{n}\right)$$

Continuing in the same way, we get

$$\zeta_2 = \bar{F}^{-1}\left(\frac{s(k-1)}{n}\right), \cdots\cdots, \zeta_k = \bar{F}^{-1}\left(\frac{s}{n}\right).$$

The feedback procedure is as follows:

1) **Threshold Determination:** BS decides on thresholding levels $\zeta_1, \zeta_2, ..., \zeta_k$ based on the sparsity level that can be recovered. For each threshold interval $[\zeta_i, \zeta_{i+1})$, repeat the *User Feedback* step.

2) **User Feedback:** Repeat the following steps for each beam.

   - CQI Determination: Each user determines his best beam (corresponding to the highest CQI value).
   - CQI Feedback: Each user feeds back his CQI if it lies in threshold interval $[\zeta_i, \zeta_{i+1})$ on all shared channels. Otherwise, the user remains silent.
   - Compressive Sensing: BS finds the strong users using Compressive Sensing.
   - Least-squares estimation/refining: BS estimates or refines results obtained via CS using least-squares.

3) **User Selection:** For each beam, BS randomly selects one of strong users of the highest active threshold interval, where active threshold interval here means that there is at least one user sending feedback data in the interval. Here, CQI is the lower limit of the highest active threshold interval.

## IV. Throughput in the RBF Case

In this section, we present the sum-rate throughput achieved by the proposed schemes. Although we focus on RBF, the proposed schemes can be applied to other beamforming methods (e.g. ZFBF).

### A. Throughput in the Analog Feedback Case

The sum-rate throughput achieved in the RBF case with dedicated ideal feedback links is given by [4]

$$\mathcal{R} \approx \mathbb{E}\left[\sum_{m=1}^{p} \log_2(1 + \max_{1 \leq i \leq n} \text{SINR}_{i,m})\right] \tag{6}$$

Also, it is shown in [4] that (6) is equivalent to

$$\mathcal{R} \approx p \log_2(1 + \rho \log(n) - \rho(p - 2) \log \log(n)) \tag{7}$$

As the SINRs fed back by the users are transmitted *as is* and the feedback links are noisy, so there is a need to back off the noisy received SINRs based on the noise variance as follows:

$$\text{SINR}' = \text{SINR} + w \tag{8}$$

where the actual and noisy SINRs are denoted by SINR and SINR$'$ respectively and $w$ represents noise. Now, if we decide to back off the received SINRs by an amount $\triangle$, then the back-off efficiency ($\eta$) i.e. the probability that this backed off SINR is less than or equal to the actual SINR is given as follows:

$$\eta = \mathbb{P}[\text{SINR}' - \triangle \leq \text{SINR}] = \mathbb{P}[w \leq \triangle] = 1 - Q\left(\frac{\triangle}{\sigma_{\mathbf{w}}}\right)$$

where $Q$ represents the Q-function. Thus, the effective throughput (with back-off on noisy SINR) can be written as:

$$\mathcal{R}_{eff} \approx \left(1 - Q\left(\frac{\triangle}{\sigma_{\mathbf{w}}}\right)\right) p \log_2(\beta - \triangle) \tag{9}$$

where $\beta = 1 + \rho \log(n) - \rho(p - 2) \log \log(n)$.

Differentiating $\mathcal{R}_{eff}$ w.r.t. $\triangle$ and setting it equal to 0, yields

$$Q\left(\frac{\triangle}{\sigma_{\mathbf{w}}}\right) + \left(\frac{\beta - \triangle}{\sqrt{2\pi}\sigma_{\mathbf{w}}}\right) \exp\left(-\frac{\triangle^2}{2\sigma_{\mathbf{w}}^2}\right) \log(\beta - \triangle) = 1 \tag{10}$$

Simulation results confirm that the value of $\triangle$ that satisfies the above equation maximizes the effective throughput.

### B. Throughput in the Digital Feedback Case

The sum-rate throughput achieved by $p$ beams in the multiple thresholds ($k$ in number) based digital feedback case for RBF is given below.

$$\mathcal{R} \approx p\mathbb{E}\left[\log_2(1 + \max_{1 \leq i \leq k} \zeta_i)\right]$$

where $\max\limits_{1 \leq i \leq k} \zeta_i$ is the lower limit of the CQI of the highest active threshold interval.

Alternatively, the same throughput can be derived analytically as follows. The throughput achieved for any transmit beam $m$ is given as follows:

$$\mathcal{R}_m = \sum_{i=1}^{k} \log_2(1 + \zeta_i) \mathbb{P}(\text{selected user in the threshold interval}) \mathbb{P}(\text{threshold interval})$$

The probability of the threshold interval (denoted as $Q_i$) is given by $\mathbb{P}(Q_i) = [F(\zeta_{i+1}) - F(\zeta_i)]$, where $F(\zeta)$ is the cumulative distribution function (CDF) of CQI (SINR) defined as: $F(\zeta) = \mathbb{P}[\text{SINR} \leq \zeta] = 1 - \frac{\exp(-\zeta/\rho)}{(1+\zeta)^{p-1}}, \zeta \geq 0$ [4]. Capitalizing on the work of [15], we calculate the probability that selected user is in threshold interval $Q_i$ as follows:

$$\mathbb{P}(\text{selected user is in } Q_i) = \sum_{j=0}^{n-1} \frac{1}{j+1} \binom{n-1}{j} \mathcal{P}_1 \mathcal{P}_2$$

where

$$\mathcal{P}_1 = \mathbb{P}(j \text{ users other than the selected user are in } Q_i) = [F(\zeta_{i+1}) - F(\zeta_i)]^j, \text{and}$$

$$\mathcal{P}_2 = \mathbb{P}((n-j-1) \text{ users lies below the interval } Q_i) = [F(\zeta_i)]^{(n-j-1)}$$

Substituting these values of $\mathcal{P}_1$ and $\mathcal{P}_2$, and after some manipulations, one can show that

$$\mathbb{P}(\text{selected user is in } Q_i) = \frac{[F(\zeta_{i+1})]^n - [F(\zeta_i)]^n}{[F(\zeta_{i+1}) - F(\zeta_i)]}$$

Thus,

$$\mathcal{R}_m = \sum_{i=1}^{k} \log_2(1 + \zeta_i)([F(\zeta_{i+1})]^n - [F(\zeta_i)]^n)$$

As, in our case there are $p$ beams and all of them are identical, so the sum-rate throughput is given as

$$\mathcal{R} = \sum_{m=1}^{p} \mathcal{R}_m = p \sum_{i=1}^{k} \log_2(1 + \zeta_i)([F(\zeta_{i+1})]^n - [F(\zeta_i)]^n)$$

## V. Performance Evaluation

We consider following metrics for the performance evaluation of the proposed feedback schemes.

### A. Feedback Resources Reduction

There is a significant reduction in number of feedback channels required for carrying feedback information. The proposed schemes requires only $O(\log(n))$ feedback channels (shown in the Lemma given below) as opposed to $n$ feedback channels required in the dedicated feedback case.

*Lemma 2: The number of multiple access feedback channels required for the proposed schemes is $\frac{c}{2}(s \log(n))$, where $c$ is a constant.*

Proof: Specifically, let's assume that there are $r$ channels shared between users over which feedback can take place. We can represent these channels using the system of equations (2). As already mentioned, (2) is similar to ones considered in [13]- [14], except that in our case the measurement matrix $\mathbf{A}$, and the noise vector $\mathbf{w}$ are complex instead of real. So, we replace the complex-valued model in (2) by its real-valued equivalent as shown below

$$\left[ \begin{array}{c} \Re(\mathbf{y}) \\ \Im(\mathbf{y}) \end{array} \right] = \left[ \begin{array}{cc} \Re(\mathbf{A}) & -\Im(\mathbf{A}) \\ \Im(\mathbf{A}) & \Re(\mathbf{A}) \end{array} \right] \left[ \begin{array}{c} \mathbf{v} \\ 0 \end{array} \right] + \left[ \begin{array}{c} \Re(\mathbf{w}) \\ \Im(\mathbf{w}) \end{array} \right]$$

where $\Re(\mathbf{A})$ & $\Im(\mathbf{A})$ represents real and imaginary part of $\mathbf{A}$. After simplification the above equation reduces to

$$\left[ \begin{array}{c} \Re(\mathbf{y}) \\ \Im(\mathbf{y}) \end{array} \right] = \left[ \begin{array}{c} \Re(\mathbf{A}) \\ \Im(\mathbf{A}) \end{array} \right] \left[ \begin{array}{c} \mathbf{v} \end{array} \right] + \left[ \begin{array}{c} \Re(\mathbf{w}) \\ \Im(\mathbf{w}) \end{array} \right].$$

or,

$$\underline{\mathbf{y}} = \underline{\mathbf{A}}\mathbf{v} + \underline{\mathbf{w}} \tag{11}$$

The entries of $\underline{\mathbf{A}}$ are i.i.d. $\mathcal{N}\left(0, 1/2\right)$, and the entries of $\underline{\mathbf{w}}$ are i.i.d. $\mathcal{N}\left(0, \sigma^2/2\right)$. The above model (V-A) gives us the $2r \times n$ real measurement matrix, and $2r \times 1$ real noise vector, so the sparsity pattern recovery techniques discussed in Section III-A can be applied. Also, note that small values of $s$ are sufficient (Section III-B). Therefore, we have

$$2r \approx cs\log(n), \quad \Rightarrow \quad r = \frac{c}{2}(s\log(n)). \tag{12}$$

*Lemma 3:* In the RBF case when $n \to \infty$, the minimum number of multiple access feedback channels required is $(\log \log \log(n)) \log(n)$.

Proof: From *Lemma 2*, we have $r = \frac{c}{2}(s\log(n))$ and for $n \to \infty$, $c = 2$ [13]. For RBF systems with large number of users ($n \to \infty$), the minimum value of $s$ (the number of users who should feedback) required to achieve the sum-rate throughput is given by $\log \log \log(n)$ [16]. Substituting these value of $c$ and $s$ in $r = \frac{c}{2}s\log(n)$, we conclude that number of multiple access (shared) feedback channels is $r = (\log \log \log(n)) \log(n)$.

## B. Feedback Noise Reduction in the Analog Feedback Case

The other important benefit of this scheme is the feedback noise reduction (which eventually results in better throughput) in the analog feedback case. This is because the feedback data of each user is carried over all shared channels. This come in contrast to dedicated channel feedback case where feedback is carried over one channel only. We analyze the error covariance matrix (ECM), as it will allow us to identify the optimum amount of back off required on the noisy SINR which depends on the noise variance. Also, we analyze two measures of ECM — trace and determinant of error covariance matrix.

*1) Shared Feedback Channels:* Error covariance matrix after the sparsity pattern is identified and LS is applied is given by [17]

$$\text{ECM} = [\mathbf{R_v}^{-1} + \mathbf{A}_S^* \mathbf{R_w}^{-1} \mathbf{A}_S]^{-1} \tag{13}$$

where $\mathbf{R_v} = \mathbb{E}[\mathbf{v}_S \mathbf{v}_S^*] = \sigma_\mathbf{v}^2 \mathbf{I}$, and $\mathbf{R_w} = \mathbb{E}[\mathbf{w}_S \mathbf{w}_S^*] = \sigma_\mathbf{w}^2 \mathbf{I}$. $\mathbf{v}_S$ and $\mathbf{w}_S$ refers to the entries of $\mathbf{v}$ and $\mathbf{w}$ corresponding to $S$. Substituting these values in (13), yields

$$\mathbb{E}[\text{ECM}] = \mathbb{E}_{\mathbf{A}_S^* \mathbf{A}_S} \left[ \left( \frac{1}{\sigma_\mathbf{v}^2} \mathbf{I} + \frac{1}{\sigma_\mathbf{w}^2} \mathbf{A}_S^* \mathbf{A}_S \right)^{-1} \right] \tag{14}$$

$$\overset{(a)}{\approx} \left( \frac{\sigma_\mathbf{w}^2}{\frac{\sigma_\mathbf{w}^2}{\sigma_\mathbf{v}^2} + r} \right) \mathbf{I} \overset{(b)}{\approx} \frac{\sigma_\mathbf{w}^2}{r} \mathbf{I} \tag{15}$$

where $(a)$ follows because for fixed $s$ and large $r$, $\mathbb{E}[\mathbf{A}_S^* \mathbf{A}_S] \to r\mathbf{I}$ [18], and $(b)$ follows because for large $r$ and high SNR $(\frac{\sigma_\mathbf{w}^2}{\sigma_\mathbf{v}^2} + r) \to r$. For $\sigma_\mathbf{v}^2 = 1$, $\sigma_\mathbf{w}^2 = 1/\rho$ where $\rho$ is SNR.

- Trace of ECM: Using (14),

$$\mathbb{E}[\text{tr(ECM)}] = \mathbb{E}_{\|\mathbf{a}_1\|^2 \cdots \|\mathbf{a}_s\|^2} \left[ \sum_{i=1}^s (1 + \rho \|\mathbf{a}_i\|^2)^{-1} \right] = s \mathbb{E}_{\|\mathbf{a}\|^2} [(1 + \rho \|\mathbf{a}\|^2)^{-1}]$$

where $\|\mathbf{a}\|^2$ is a chi-squared variable with $2r$ degrees of freedom. Therefore,

$$\mathbb{E}[\text{tr(ECM)}] = s \int_0^\infty (1 + \rho x)^{-1} \frac{1}{\Gamma(r)} x^{(r-1)} e^{-x} dx$$

where $\Gamma(\cdot)$ is the gamma function [19]. Using (3.383.10) of [19], the above evaluates to

$$\mathbb{E}[\text{tr(ECM)}] = s(1/\rho)^r e^{1/\rho} \Gamma(1 - r, 1/\rho)$$

where $\Gamma(\cdot, \cdot)$ is the incomplete gamma function [19].

- Determinant of ECM: Using (14),

$$\mathbb{E}[\text{det(ECM)}] = \mathbb{E}_{\mathbf{A}_S^* \mathbf{A}_S} \left[ \det \left\{ \left( \frac{1}{\sigma_\mathbf{v}^2} \mathbf{I} + \frac{1}{\sigma_\mathbf{w}^2} \mathbf{A}_S^* \mathbf{A}_S \right)^{-1} \right\} \right]$$

We can write the above equation in terms of eigenvalues of $\mathbf{A}_S^* \mathbf{A}_S$ as follows

$$\mathbb{E}[\det(\text{ECM})] = \mathbb{E}_{\lambda_1 \cdots \lambda_s} \left[ \prod_{i=1}^{s} \left( \frac{1}{\sigma_{\mathbf{v}}^2} + \frac{\lambda_i}{\sigma_{\mathbf{w}}^2} \right)^{-1} \right]$$

$$= \prod_{i=1}^{s} \mathbb{E}_{\lambda} \left[ \left( \frac{1}{\sigma_{\mathbf{v}}^2} + \frac{\lambda}{\sigma_{\mathbf{w}}^2} \right)^{-1} \right]$$

$$= \left[ \int_0^{\infty} \left( \frac{1}{\sigma_{\mathbf{v}}^2} + \frac{\lambda}{\sigma_{\mathbf{w}}^2} \right)^{-1} p(\lambda) d\lambda \right]^s.$$

$p(\lambda)$ is given by $\frac{1}{s} \sum_{l=0}^{s-1} \frac{l!}{(l+r-s)!} [L_l^{r-s}(\lambda)]^2 \lambda^{r-s} e^{-\lambda}$ [18], where $L_l^{r-s}(\lambda) = \frac{1}{l!} e^{\lambda} \lambda^{s-r} \frac{d^l}{d\lambda^l} (e^{-\lambda} \lambda^{r-s+l})$ is the associated Laguerre polynomial of order $l$.

Alternatively, if we use the approximation in (15), trace and determinant of ECM are given by $\left( \frac{s}{\rho r} \right)$ and $\left( \frac{1}{\rho r} \right)^s$ respectively.

*2) Dedicated Feedback Channels:*

$$\text{ECM} = \left[ \left( \frac{1}{\sigma_{\mathbf{v}}^2} \mathbf{I} + \frac{1}{\sigma_{\mathbf{w}}^2} \mathbf{I} \right)^{-1} \right] = \left( \frac{\sigma_{\mathbf{w}}^2}{\frac{\sigma_{\mathbf{w}}^2}{\sigma_{\mathbf{v}}^2} + 1} \right) \mathbf{I} \quad (16)$$

For $\sigma_{\mathbf{v}}^2 = 1$, $\sigma_{\mathbf{w}}^2 = 1/\rho$ where $\rho$ is SNR.

- Trace of ECM: Using (16), $\text{tr}(\text{ECM}) = n \left( 1 + \rho \right)^{-1}$.
- Determinant of ECM: Using (16), $\det(\text{ECM}) = \left( 1 + \rho \right)^{-n}$.

*3) Comparison between Shared & Dedicated Feedback Channel Cases:* Trace of ECM and its approximation is plotted in Fig. 1. Similar plot for determinant is omitted due to space limitation. Thus, from (15) & (16), we conclude that the back off on the SINR is $O(\frac{\sigma_{\mathbf{w}}}{\sqrt{r}})$ in the shared feedback channels case as opposed to $O(\sigma_{\mathbf{w}})$ in the dedicated feedback channel case.

Another point that needs to be noted is that the trace of ECM (which is commonly refereed as cost function and should be minimized [17]) in the shared feedback channel case is much smaller than that obtained in the dedicated feedback channel case.

## C. Feedback Load Reduction

In addition to the feedback resources reduction, there is a reduction in the amount of feedback. In RBF scheme with dedicated feedback channel [4], $n$ real values and $n$ integer values ($n \log_2 p$ bits) are fedback, as there are $n$ users in the system.

*1) Analog Feedback Case:* The proposed CS based analog feedback scheme requires only $pr$ real values to be fedback. This is because there are $r$ shared channels and the scheme is repeated for each beam. Note that the feedback load reduction is more dominant in systems with large number of users, as $r \sim O(\log(n))$ and $p$ is small.

*2) Digital Feedback Case:* The proposed CS based digital feedback scheme requires only $pkr$ bits to be fedback. This is because there are $r$ shared channels and the scheme is repeated for each beam & threshold. Note that the feedback load reduction is more dominant in systems with large number of users, as $r \sim O(\log(n))$ and $p$ & $k$ are small.

### D. Trade-off in the Digital Feedback Case

Given a budget of bits that can be fedback, using intuition, it was shown in [20] that trade-off exists between the multi-user diversity and feedback accuracy. In our context, multi-user diversity is related to the number of shared channels $r$ whereas feedback accuracy is related to the number of thresholds $k$, and so a similar trade-off may exist. The number of shared channels and thresholds must be chosen such that the throughput is maximized. This is explored using simulation in section VII.

## VI. Feedback Channel Training

In the previous sections, we assumed that the channel $\mathbf{A}$ estimation is given to the system with the aid of a "genie" at no cost. In this section, we present how the feedback channel training can be accomplished and explore ways to reduce it. Here, we assume that (2) represents frequency feedback channels i.e., the entries of $\mathbf{A}$, $a_{ij}$ represents the gain of the $j$-th user in the $i$-th frequency band.

### A. Channel Matrix is Full

The optimal number of symbols required for channel training is equal to the number of transmit antennas [21]. So we need $p$ training symbols for the downlink channel and $n$ training symbol for the uplink channel (as there are $n$ users each having one transmit antenna). Training for each user in the uplink can be performed one by one, i.e., the first symbol of the coherence interval is reserved for user $1$ to perform training for all shared channels, and second symbol reserved for user $2$, and so on. Continuing in this way, we need $n$ symbol time to accomplish training for all users. Also, it is important to note that as there is little data to be sent for feedback purposes, so much of the uplink coherence time can be used for feedback training. Coherence time is typically of the order of few thousand symbols, so training would not be an issue for systems with moderate number of users. However, a method for reducing the amount of feedback channel training time is discussed in the next subsection.

### B. Channel Matrix is Block Diagonal

In order to reduce the feedback training time, we divide the users into groups with each group being allowed to feedback only on a set of feedback channels, thereby reducing the full channel matrix to a

block diagonal one

$$\mathbf{A}_{BD} = \begin{bmatrix} \mathbf{A}_1 & & \\ & \ddots & \\ & & \mathbf{A}_k \end{bmatrix}.$$

Compressive sensing is applied in the same way as discussed in Section III, the only difference being that it is now applied on each block. Strong users in each block (or group) are found and the user corresponding to the maximum SINR among the strong users from all blocks is selected. As the users are i.i.d., so we divide the feedback resource equally among the $k$ groups. Thus, training can now be performed for each block simultaneously. This approach reduces the feedback training time considerably, e.g. if we divide the total number of users into two groups, then the training will require $n/2$ symbol time as opposed to $n$ symbol time required for the case when the channel matrix is full.

The flip side of this approach is that compressive sensing is now applied on the group of users instead of all users as one block. Thus, for same sparsity level $s$ (overall), with block diagonalization, the number of feedback channels required is given below

$$f_{\mathbf{A}_{BD}} = f_{\mathbf{A}_1} + \cdots + f_{\mathbf{A}_k}$$
$$= k f_{\mathbf{A}_1}$$
$$= k \left[ \frac{1}{2} \left( c' \left( \frac{s}{k} \right) \log \left( \frac{n}{k} \right) \right) \right].$$

Note that from the above equation it may first appear that the number of channels have reduced as the quantity inside the logarithm is reduced by a factor of $k$, however, it is the other way round. This is because now $c'$ has increased as the problem dimension ($n$) is reduced by a factor of $k$ [13]. Thus, there is a trade-off between the reduction in the amount of feedback training and the number of feedback channels. Also, note that there is now an additional constraint requiring $s/k$ to be an integer.

### C. Non-fading Channels

When the channels are non-fading i.e., the channel gains are constant (or 1), then each strong users multiples its CQI with a unique binary chip sequence (consisting of $\pm 1$ each with probability 0.5) of length equal to the number of shared feedback channels $r$ and send it over the multiple access shared channels. There are two ways of assigning chip sequences to the users: pre-programmed in users' device or sending it over the air. If it is send over the air, then the training time (used in the case of fading channels) can be used to send unique binary chip sequences to all $n$ users. Thus, $\mathbf{A}$ is $r \times n$ Bernoulli matrix and so CS can be applied as Bernoulli matrices are shown to satisfy the RIP [10].

## VII. Numerical Results

In this section, we present numerical results for CS-based feedback schemes by applying it in RBF context. We use $p = 4$ base station antennas, and $n = 100$ users. We set the threshold according to the sparsity level $s$, and use the maximum correlation technique (unless mentioned otherwise) for compressive sensing as this is much more computationally efficient than LASSO. Each point in the figures represents the sum-rate throughput achieved for shared number of channels determined by $c$ & $s$ according to (12). We use SNR = 10 dB for both downlink and feedback link (unless stated otherwise) for calculating the sum-rate throughput.

### A. Analog Feedback Case

In Fig. 2, we present the sum-rate throughput with shared channel feedback in the analog feedback case. We use optimum back off on noisy SINRs in the analog feedback case. From this figure, we note that for small values of $s$ the throughput is low. This is because the threshold works well for systems with large number of users but for systems with moderate number of users, we may have more or less number of users above the threshold than desired. So, if we set $s$ low, then the probability that a beam has no strong user is relatively higher (resulting in a multiplexing loss) to the case when $s$ is large. However, large values of $s$ requires more feedback channels. Also, we see that the number of shared channels required to achieve the maximum possible throughput obtained in a noisy dedicated feedback scenario is 11 (corresponds to $c/2 = 0.4$ and $s = 6$). Also, it worth mentioning that the proposed scheme comes close to achieving the throughput obtained in a noiseless dedicated feedback scenario (dedicated feedback with ideal feedback links) due to feedback noise reduction. Note that $90\%$ of throughput in noiseless dedicated feedback case is achieved by 19 shared channels (corresponds to $c/2 = 0.8$ and $s = 5$).

In Fig. 3, we present results on block diagonalization method proposed for reducing the feedback training time (section VI-B). Here, we divide $100$ users into two groups of $50$ users and compressive sensing is applied on each group. It is clear from the figure that this method requires few more feedback channels. Also in this figure, we present result based on LASSO which shows that LASSO method performs marginally better than maximum correlation method.

In Fig. 4, we present the sum-rate throughput achieved by two-stage RBF in the analog feedback case when the feedback channels are noiseless. In the second-stage of the two-stage RBF, additional feedback information (beam gain information (BGI) [6]) is requested from the selected users only and Iterative Beam Power Control (IBPC) algorithm proposed in [6] is used for the re-distribution of the total power among the active beams (beams for which there are strong users) in an optimized manner. From the figure, we see that there is hardly any gain for two stage RBF with dedicated channel feedback, however, it is evident that for compressive sensing based opportunistic feedback protocol, two-stage RBF is effective even for moderate to large number of users. This is because if no user is strong for some beams, the system still suffers from the multiplexing loss but the power of those beam are distributed among the active beams in an optimized way. Also, note that there is no back-off required here as the feedback links are noiseless. With two-stage RBF, the number of shared feedback channels required is 15 (corresponds to $c/2 = 0.8$ and $s = 4$).

### B. Digital Feedback Case

For all digital feedback cases, we chose $s = 1$ (the minimum possible value) and set multiple thresholds as discussed in section III-D. This is because for the proposed scheme, $s = 1$ will allow us to set the highest possible uppermost threshold thereby ensuring a higher throughput.

In Fig. 5, we present the sum-rate throughput achieved with shared channel feedback in the digital feedback case. Is is evident form the figure that the proposed scheme in a noisy scenario achieves the throughput obtained in a noiseless dedicated feedback scenario (dedicated feedback with ideal feedback links). Also, we see that the throughput increases with the increase in the number of shared channels & thresholds. Taking the pessimistic view, we need only 10 feedback channels (corresponds to $c/2 = 2$ and $s = 1$). However, it is important to note that beyond a certain number of shared channels or thresholds, the throughput either becomes stagnant or increases marginally.

In Fig. 6, we consider fixed budgets of $p \times kr$ bits that can be fedback. From the figure, we note that such a trade-off exists and for a given fixed budget there is an optimum number of thresholds and shared feedback channels that maximizes the throughput.

### VIII. CONCLUSIONS

In this paper, a generic feedback channel model and compressive sensing based opportunistic feedback schemes are proposed. The proposed generic feedback channel model is shown to encompass all existing feedback channel models proposed in the literature. We have shown that the proposed analog & digital opportunistic feedback schemes achieves the same sum-rate throughput as that achieved by dedicated feedback schemes, but with feedback channels growing only logarithmically with number of users. Also, we derived an expression for the sum-rate throughput in the digital feedback case with multiple thresholds.

In the analog feedback case (noisy scenario), it has also been shown that due to feedback noise reduction, the proposed scheme comes close to achieving the throughput obtained in the case of noiseless dedicated feedback. In the digital feedback case, it has also been shown that beyond a certain number of shared channels or thresholds, the throughput either becomes stagnant or increases marginally. Also, given a budget on the amount of bits that can be fedback, we have shown that there exist a trade-off between the number of shared channels and thresholds and therefore they must be chosen such that the throughput is maximized.

Although the results presented here only show the performance of the the proposed schemes in the RBF context, the schemes can easily work with other beamforming methods.

## IX. APPENDIX

### COMPRESSIVE SENSING

Here, we give the reader a brief introduction about compressive sensing. Let $\mathbf{v} \in \mathbb{R}^n$ be an unknown vector, with at most $s$ non-zero entries ($s \leq n/2$) and let $S$ denote its support set with $|S| = s << n$. Suppose that we make a set $\{y_1, \ldots, y_r\}$ of $r$ independent and identically distributed (i.i.d.) observations of the unknown vector $\mathbf{v}$, each of the form

$$y_i = \mathbf{a}_i^T \mathbf{v} + w_i \tag{17}$$

where $w_i \sim \mathcal{N}(0, N_o)$ is observation noise, and $\mathbf{a}_i \sim \mathcal{N}(0, \mathbf{I}_{n \times n})$ is a measurement vector. In the matrix form, it can be compactly written as

$$\mathbf{y} = \mathbf{A}\mathbf{v} + \mathbf{w} \tag{18}$$

Reconstruction will not be possible if the measurement process damages the information in $\mathbf{v}$, which often happens in practice. A necessary and sufficient condition for the system of equations to be well-conditioned (thus having a stable inverse) is the *restricted isometry property* (RIP) [10]- [12].

*Definition 1*: A $r \times n$ matrix $\mathbf{A}$ has the $s$-RIP with appropriately chosen constant $0 \leq \epsilon_s < 1$ if

$$1 - \epsilon_s \leq \frac{\|\mathbf{A}\mathbf{v}\|_2^2}{\|\mathbf{v}\|_2^2} \leq 1 + \epsilon_s \tag{19}$$

holds for all $s$-sparse vectors $\mathbf{v}$.

The $s$-RIP property ensures that the matrix $\mathbf{A}$ preserves the lengths of these particular $s$-sparse vectors. Practical recovery algorithms require that $\mathbf{A}$ satisfies a more conservative RIP ($3s$-RIP in general [12]). If $\mathbf{A}$ is a random matrix consisting of Gaussian random variables, the RIP property is satisfied with overwhelming probability [11]. In this case, the number of measurements that are necessary to recover $\mathbf{v}$ efficiently in a noiseless scenario with high probability is on the order of $r \sim s \log(n/s)$.

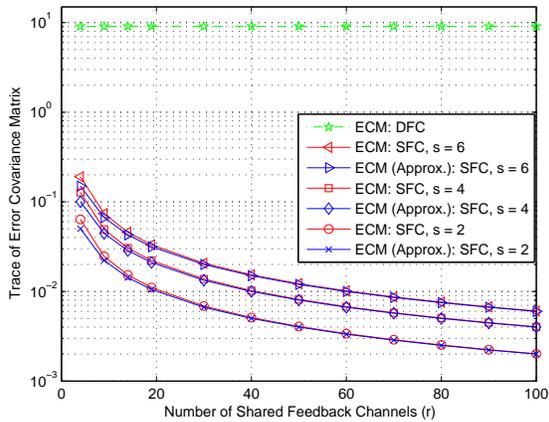

Figure 1. Trace of Error Covariance Matrix for shared feedback channel (SFC) and dedicated feedback channel (DFC), $n = 100$ and uplink SNR = 10 dB for different values of $s$.

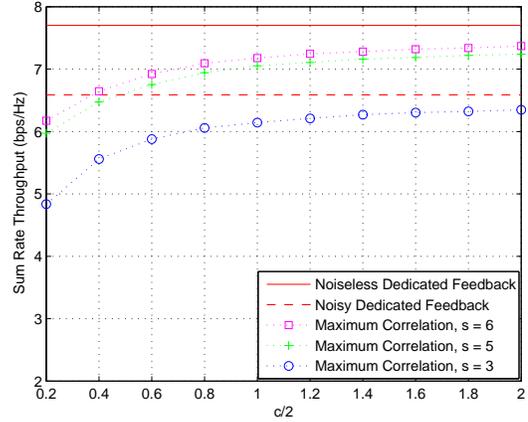

Figure 2. Analog Shared Channel Feedback: Throughput versus c/2 for RBF, $p = 4$, $n = 100$ and SNR = 10 dB (both downlink & feedback link) for different values of $s$.

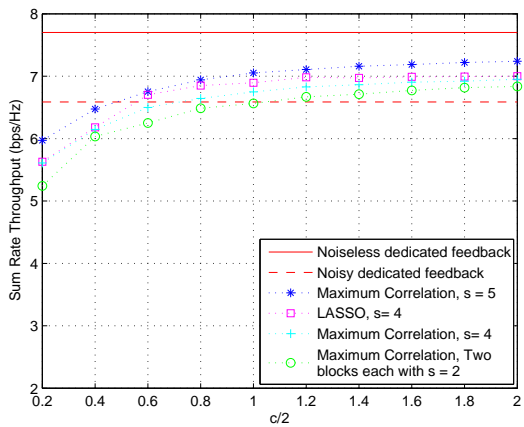

Figure 3. Analog Shared Channel Feedback: Throughput versus c/2 for RBF, $p = 4$, $n = 100$ and SNR = 10 dB (both downlink & feedback link) for different methods viz. LASSO, maximum correlation, and maximum correlation with block diagonalization.

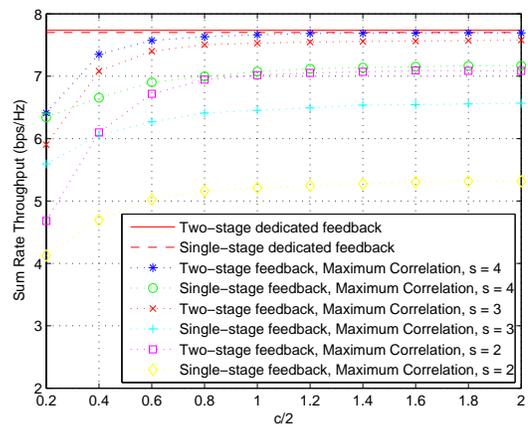

Figure 4. Analog Shared Channel Feedback: Throughput versus c/2 for Single stage and Two stage RBF, $p = 4$, $n = 100$ and SNR = 10 dB & $\infty$ for downlink and feedback link respectively.

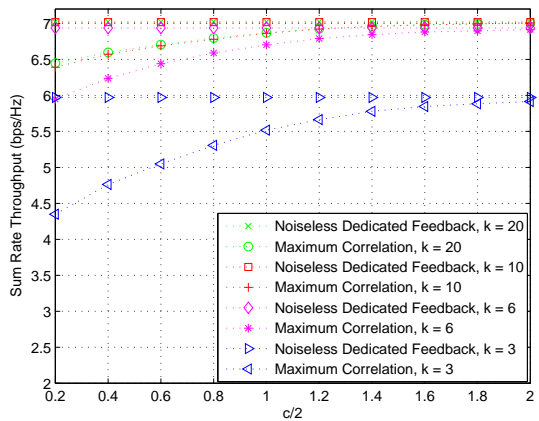

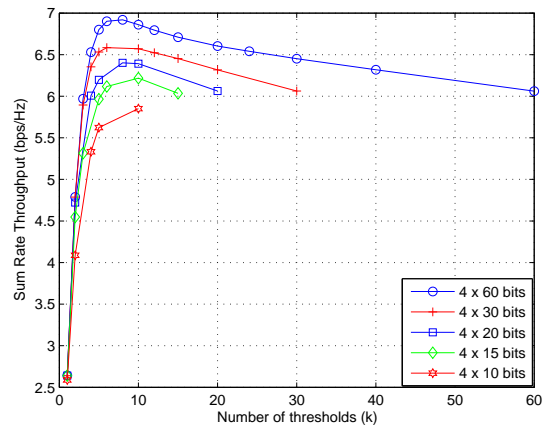

Figure 5. Digital Shared Channel Feedback: Throughput versus c/2 for RBF, $p = 4$, $n = 100$ and SNR = 10 dB (both downlink & feedback link) for different values of $k$.

Figure 6. Digital Shared Channel Feedback: Throughput versus k for RBF, $p = 4$, $n = 100$ and SNR = 10 dB (both downlink & feedback link) for different budgets of bits that are to be fedback.

| Reference | Feedback Protocol | Entries of A | Noise variance ($\sigma^2$) | BF Type | Feedback Components |
|---|---|---|---|---|---|
| Sharif et. al. [4] | Dedicated | const. | 0 | RBF | BI & SINR |
| Yoo et. al. [5] | Dedicated | const. | 0 | ZFBF | QCI & SNR/SINR |
| Kountouris et al. [6] | Dedicated | const. | 0 | RBF (1st stage) | BI & SINR |
|  |  |  |  | RBF (2nd stage) | BGI |
| Kountouris et al. [6] | Dedicated | const. | 0 | ZFBF | QCI & SINR |
| Diaz et. al. [7] | Dedicated | const. | 0 | RBF | 1 bit |
| Tang et. al. [8] | Opportunistic | const. | 0 | SISO case | ID |
| Rajiv et. al. [9] | Opportunistic | const. | 0 | RBF | ID |
|  |  |  |  | ZFBF | ID & QCI |
| Proposed | Opportunistic | $\mathcal{CN}(0,1)$ | $> 0$ | RBF | CQI (Analog Case) |
|  |  |  |  | RBF | 1 bit (Digital Case) |

Table I

GENERIC FEEDBACK CHANNEL MODEL